\documentclass[twocolumn,pra,amsmath,amssymb,floatfix,superscriptaddress]{revtex4-1}
\usepackage{color}
\usepackage{graphicx}
\usepackage{dcolumn}
\usepackage{bm}

\begin{document}


\title{Thermal evolution of antiferromagnetic correlations and tetrahedral bond angles in superconducting FeTe$_{1-x}$Se$_x$}

\author{Zhijun~Xu}
\affiliation{Physics Department, University of California, Berkeley, California 94720, USA}
\affiliation{Materials Science Division, Lawrence Berkeley National Laboratory, Berkeley, California 94720, USA}
\author{J. A. Schneeloch}
\affiliation{Condensed Matter Physics and Materials Science
Department, Brookhaven National Laboratory, Upton, New York 11973,
USA}
\author{Jinsheng~Wen}
\affiliation{Center for Superconducting Physics and Materials, National Laboratory of Solid State Microstructures and Department of Physics, Nanjing University, Nanjing 210093, China}
\author{E. S. Bo\v{z}in}
\affiliation{Condensed Matter Physics and Materials Science Department, Brookhaven National Laboratory, Upton, New York 11973, USA}
\author{G. E. Granroth}
\affiliation{Neutron Data Analysis and Visualization Division, Oak Ridge National Laboratory, Oak Ridge, TN, 37831, USA}
\author{B. L. Winn}
\affiliation{Quantum Condensed Matter Division, Oak Ridge National Laboratory, Oak Ridge, TN 37831, USA}
\author{M. Feygenson}
\affiliation{Chemical and Engineering Materials Division, Oak Ridge National Laboratory, Oak Ridge, TN 37831, USA}
\author{R. J. Birgeneau}
\affiliation{Physics Department, University of California, Berkeley, California 94720, USA}
\affiliation{Materials Science Division, Lawrence Berkeley National Laboratory, Berkeley, California 94720, USA}
\author{Genda~Gu}
\author{I. A. Zaliznyak}
\author{J.~M.~Tranquada}
\author{Guangyong~Xu}
\affiliation{Condensed Matter Physics and Materials Science
Department, Brookhaven National Laboratory, Upton, New York 11973,
USA}
\date{\today}

\begin{abstract}
It has recently been demonstrated that dynamical magnetic correlations measured by neutron scattering in iron chalcogenides can be described with models of short-range correlations characterized by particular  {choices of four-spin plaquettes, where the appropriate choice changes as the} parent material is doped towards superconductivity. Here we apply such models to describe measured maps of magnetic scattering as a function of two-dimensional wave vectors obtained for optimally superconducting crystals of FeTe$_{1-x}$Se$_x$. We show that the characteristic antiferromagnetic wave vector evolves from that of the bicollinear structure found in underdoped chalcogenides (at high temperature) to that associated with the stripe structure of antiferromagnetic iron arsenides (at low temperature); {these can both be described with the same local plaquette, but with different inter-plaquette correlations}. While the magnitude of the low-energy magnetic spectral weight is substantial at all temperatures, it actually weakens somewhat at low temperature, where the charge carriers become more itinerant. The observed change in spin correlations is correlated with the dramatic drop in the electronic scattering rate and the growth of the bulk nematic response on cooling. Finally, we also present powder neutron diffraction results for lattice parameters in FeTe$_{1-x}$Se$_x$ indicating that the tetrahedral bond angle tends to increase towards the ideal value on cooling, in agreement with the increased screening of the crystal field by more itinerant electrons and the correspondingly smaller splitting of the Fe $3d$ orbitals.
\end{abstract}


\maketitle

\section{Introduction}

The roles of magnetic fluctuations and orbital ordering are at the center of a continuing debate in the field of iron-based superconductors (FeBS).   While their contributions to the superconducting mechanism are of particular interest \cite{Mazin2008,Kuroki2008,Kontani2010}, another forum concerns the nature of the nematic electronic response \cite{Fernandes2014}. In an attempt to look for minimal models, the discussion is often focused on an exclusive choice: either magnetic correlations \cite{Fang2008,Xu2008,Mazin2009} or orbital fluctuations \cite{Lv2009,Kruger2009} are the dominant factor.

Experimental evidence for nematic response was first obtained in the BaFe$_2$As$_2$ system \cite{Chu2010,Yi2011,Chu2012,Kuo2013,Shapiro2015,Lu2014}, where a structural transition that lowers the rotational symmetry from $C_4$ to $C_2$ is closely followed by antiferromagnetic ordering \cite{Huang2008,Chu2009,Pratt2009}, with modulation wave vector $(\pi,0)$ \footnote{Here we have used the notation commonly used by theorists, which assumes a single Fe per unit cell and units of $1/a_0$.  When discussing experimental data, we will refer to a unit cell containing two Fe atoms, in units of $2\pi/a$.}.  New interest has been generated by the recent observations of nematicity in FeSe \cite{Baek2015}, a superconducting compound that exhibits a symmetry-lowering structural transition but no magnetic order \cite{Bohmer2015,Imai2009}.  Of particular interest is the observation of a temperature dependent splitting of $d_{xz}$ and $d_{yz}$ orbitals through angle-resolved photoemission spectroscopic (ARPES) studies \cite{Shimojima2014,Nakayama2014,Zhang2015}. 

While several analyses have shown that it is possible to have a nematic response due to dynamic magnetic correlations alone \cite{Wang2015,Yu2015,Glasbrenner2015}, {driving a transition to an orthorhombic phase \cite{Bohmer2015} with fluctuations alone is another matter. In any case,} one might wonder to what extent the distinction between magnetic and orbital correlations is artificial.  Experimentally, there is no question that there are substantial instantaneous magnetic moments on Fe sites in the FeBS compounds, both from x-ray emission spectroscopy \cite{Gretarsson2011,Mannella2014} and from neutron scattering \cite{Dai2015,Tranquada2014}, and that these moments are generally antiferromagnetically correlated \cite{Dai2015,Tranquada2014}, regardless of whether static order is observed.  In particular, low-energy magnetic excitations about the $(\pi,0)$ wave vector have been observed in FeSe by neutron scattering \cite{Rahn2015,WangQ2015}.  These observations are supported by theoretical calculations using Dynamical Mean Field Theory (DMFT) \cite{Yin2011}.  At the same time, a number of analyses have found that models consistent with the magnetic order also exhibit partial orbital ordering \cite{Kruger2009,Lee2009theory,Lv2009,Dai2012,Bascones2010}, involving broken degeneracy of the $d_{xz}$ and $d_{yz}$ orbitals. Indeed, an energy splitting between bands of dominant $d_{xz}$ and $d_{yz}$ character has been observed \cite{Yi2011} over the same range of temperatures as anisotropies of the intensity of spin excitations \cite{Lu2014}.

\begin{figure}[tb]
\includegraphics[width=\linewidth]{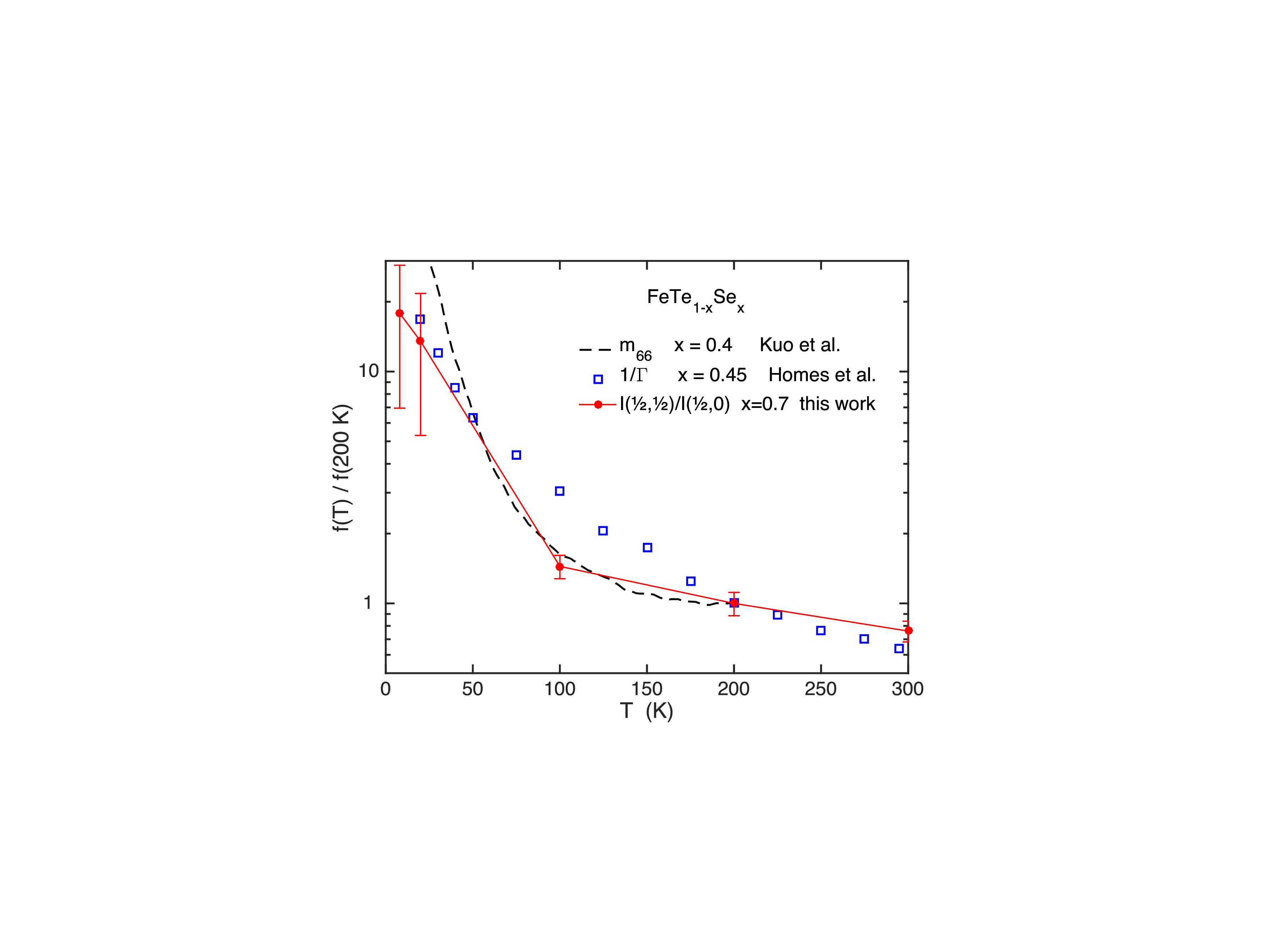}
\caption{(Color online) Comparison of the elastoresistance coefficient $m_{66}$ for $x=0.4$ (dashed line) \cite{Kuo2015}, inverse scattering rate of the narrow Drude component from optical conductivity measurements of $x=0.45$ (open squares) \cite{Homes2015}, and the ratio of 7-meV magnetic spectral weight integrated about the spin-stripe wave vector $(\frac12,\frac12)$ and the double-stripe wave vector $(\frac12,0)$ (filled circles connected by solid line), taken from Fig.~\ref{fig:4}, with error bars reflecting counting statistics.  All quantities have been normalized at 200~K.
} \label{fig:comp}
\end{figure}

While FeSe has garnered a lot of attention, FeTe$_{1-x}$Se$_x$ in the regime of optimal superconducting transition temperature ($T_c$) is also rather interesting.   Although the average crystal structure remains tetragonal, {elastoresistance measurements demonstrate a strong nematic response in the $B_{2g}$ symmetry channel that appears to diverge at low temperature, similar to other optimally-doped Fe-pnictide superconductors} \cite{Kuo2015}, as reproduced in Fig.~\ref{fig:comp}.  Furthermore, there is an evidence of the local C$_4$ symmetry breaking down to C$_2$ in the pattern of short-range dynamical magnetic correlations measured by neutron scattering in parent material FeTe with S, or Se doping, and also on cooling in a composition that is superconducting~\cite{Zaliznyak2015}, while ARPES measurements at 25~K indicate a splitting of the $xz$ and $yz$ bands at zone center \cite{Johnson2015}. The coherence of the charge carriers also shows a strong temperature dependence: optical conductivity measurements find a component that becomes coherent only at low temperature, with the inverse of the energy width growing on cooling \cite{Homes2015}.

An important aspect of the orbital nature of the electronic band structure involves splitting between bands with ${xy}$ and ${xz}/{yz}$ character, which is sensitive to deviations from the ideal tetrahedral bond angle of 109.5$^\circ$ \cite{Yin2011}. In FeTe$_{1-x}$Se$_x$, the bond angles are essentially controlled by the height of the chalcogenide ions above and below the Fe layers, with the bond angle dropping from 104$^\circ$ in FeSe to $\sim94^\circ$ in Fe$_{1+y}$Te \cite{McQueen2009,Louca2010}.  It was argued \cite{Moon2010} that the change in height of the chalcogenide ions modifies the relative magnitudes of superexchange couplings {(although the concept of superexchange is not well defined for multi-orbital FeBS with a high degree of itinerancy \cite{Mazin10})}, resulting in a change of the characteristic antiferromagnetic wave vector from that describing the double-stripe $(\pi/2,\pi/2)$ order known to occur in Fe$_{1+y}$Te \cite{Baow2009prl,Lis2009,Wen2009}, to that of the dynamical single-stripe $(\pi,0)$ correlations in superconducting FeTe$_{1-x}$Se$_x$ \cite{Qiu2009,Lee2010,Lumsden2010nf,liu10}.  {(See Table~\ref{tab:Q} for a clear definition of the wave-vector notation used in this paper.)}

In this paper, we study the temperature-dependent change of the antiferromagnetic correlations in FeTe$_{1-x}$Se$_x$ by inelastic neutron scattering.  We \cite{Xu2012PRL,Zhijun2014} and others \cite{Tsyrulin2012} have previously shown that the low-energy excitations centered at $(\pi,0)$ in the superconducting state shift in reciprocal space on warming to 100~K and above; one-dimensional cuts through the $(\pi,0)$ ``resonance'' position reveal a change from a broad commensurate peak at $(\pi,0)$ to incommensurate correlations peaked near $(0.25\pi,0.75\pi)$ and $(0.75\pi,0.25\pi)$. Here we present measurements covering two-dimensional slices of reciprocal space, finding that the main locus of the low-energy spectral weight in fact shifts from $(\pi,0)$ to $(\pi/2,\pi/2)$ on warming. Moreover, in a sample that is non-superconducting due to excess Fe, we show that the magnetic correlations remain pinned at $(\pi/2,\pi/2)$. 

{The pattern observed here at high temperature and in the non-superconducting sample is quite similar to that recently reported by one of us (IZ) \cite{Zaliznyak2015} in a S-doped FeTe sample with filamentary superconductivity.  There the pattern emerged upon cooling, replacing the high-temperature pattern characteristic of the parent Fe$_{1+y}$Te \cite{Zaliznyak2011}, indicating a transition between two different spin-liquid states.  The measured spin-spin correlations were described by a model in which a long-range spin pattern is broken into four-spin plaquettes, with exponential decay of correlations between plaquettes with distance from the central plaquette. The change in symmetry of the model plaquettes needed to simulate the measured inelastic diffuse scattering suggested local breaking of the C$_4$ symmetry down to C$_2$ on cooling, prior to reaching the superconducting state.  In the present case, we find that the spin correlations at both high and low temperatures can be modeled with the same choice of plaquette (having only C$_2$ symmetry), but with the inter-plaquette correlations changing from the double-stripe to the single-stripe wave vector on cooling.}

\begin{table}[t]
\caption{\label{tab:Q} Definitions of stripe antiferromagnetic (SAF) and double stripe antiferromagnetic (DSAF) wave vectors for two choices of unit cell.  For the 1-Fe (2-Fe) unit cell, the units are $1/a_0$ ($2\pi/a$), where $a_0$ and $a$ are the corresponding lattice parameters.}
\begin{ruledtabular}
\begin{tabular}{lcc}
  & 1-Fe unit cell & 2-Fe unit cell \\
\hline
 ${\bf Q}_{\rm SAF}$ \rule{0pt}{12pt} & $(\pi,0)$ & $(\frac12,\frac12)$ \\
 ${\bf Q}_{\rm DSAF}$ \rule{0pt}{12pt} & $(\frac{\pi}2,\frac{\pi}2)$ & $(\frac12,0)$ \\
\end{tabular}
\end{ruledtabular}
\end{table}

Besides the change in characteristic wave vector, we also observe a  decrease in low-energy magnetic weight on cooling, which parallels the increased itinerancy of charge carriers.   Such behavior is qualitatively consistent with recent theoretical work \cite{Tam2015} and previous experimental results \cite{Zaliznyak2011,Zaliznyak2015}.

To gain further insight into this curious thermal evolution, we have used neutron powder diffraction to measure the temperature dependence of the lattice parameters for a series of FeTe$_{1-x}$Se$_x$ with $0\le x\le 1$.  For $x$ away from the limiting values, we find an anomalous rise in the $a/c$ ratio on cooling, corresponding to an increase in the tetrahedral bond angles.  We infer an associated change in both the crystal-field splitting and the hybridization 
of the $t_{2g}$ states.  This, together with the evidence for nematic correlations \cite{Kuo2015} and $xz/yz$ splitting at low temperature \cite{Johnson2015}, indicates that the change in magnetic correlations with temperature can be associated with changing exchange couplings. 

The rest of the paper is organized as follows.  The experimental methods are described in the next section.  The results and analysis are presented in Sec.~\ref{sec:raa}.   The results are summarized and discussed in Sec.~\ref{sec:disc}.

\section{Experimental Methods}

The single crystals of FeTe$_{1-x}$Se$_x$ studied here were grown by unidirectional solidification \cite{Wen2011}.   Here we study superconducting samples with $x=0.50$ (SC50) and
$x=0.70$ (SC70), each with $T_c\gtrsim14$~K, and a non-superconducting sample with $x=0.45$ (NSC45) and excess Fe.  Previous characterizations of these crystals have been reported in \cite{Wen2010H,Lee2010,Zhijun2010,Zhijun_2011,Zhijun2014}.  From here on, we will specify momentum transfer ${\bf Q}=(h,k,l)$ in units of $(2\pi/a,2\pi/a,2\pi/c)$, where we assume a tetragonal unit cell with 2 Fe atoms per unit cell, as illustrated in Fig.~\ref{fig:1}(a).  (Room temperature lattice parameters are presented in Fig.~\ref{fig:acx}.)


The inelastic neutron scattering experiments were performed on time-of-flight instruments at the Spallation Neutron Source (SNS), Oak Ridge National Laboratory (ORNL).  The SC50 sample was measured on SEQUOIA (BL-17) \cite{sequoia10}, with an incident energy $E_i=40$~meV, using Fermi chopper number 1 at 360~Hz, and the [001] direction of the crystal aligned with the incident beam direction.  As the measurements were done with a fixed orientation of the crystal, we obtained data as a function of excitation energy $\hbar\omega$ and ${\bf Q} = (h,k,l_0)$, where $l_0$ is determined by $E_i$ and $\hbar\omega$.

The SC70 and NSC45 samples were measured on HYSPEC (BL-14B) \cite{hyspec15} with $E_i=20$~meV and a chopper {frequency} of 180~Hz.  With the [001] direction of the sample vertical (perpendicular to the scattering plane), the in-plane orientation was stepped by increments of 2$^\circ$ over a range of 180$^\circ$. The detectors were positioned to cover neutrons with scattering angles from 5$^\circ$ to 65$^\circ$. From the combined data, it is possible to extract slices at constant $\hbar\omega$ for the $(h,k,0)$ plane. The data have been symmetrized to enforce the 4-fold symmetry for better presentations.


The neutron powder diffraction measurements were performed at the intermediate-resolution diffractometer NOMAD (BL-1B) at the SNS \cite{Neuefeind2012}.  The samples were prepared by grinding pieces of single crystals.   The compositions correspond to $x=0,0.1,0.2,0.3,0.45,0.5,0.7,1$, where the $x=0$ sample has 10\%\ excess of Fe, while the Fe excess in the other samples has been kept close to zero; superconducting transition temperatures for all samples except for $x=1$ were reported in \cite{Katayama2010}. Note that the $x=0.50$ and $0.70$ powder samples are very similar to SC50 and SC70 samples, the $x=0.45$ powder sample does not have excess Fe (unlike NSC45) and is superconducting. Each sample of $\sim8$ g was loaded into a vanadium can under a helium atmosphere, and then cooled in an ``orange'' (liquid helium) cryostat.  Temperature-dependent measurements were performed while warming from 10~K to 300~K, with a counting time of 1 h per temperature.

To extract lattice parameters from the data, Le Bail refinements \cite{LeBail2005} were performed using GSAS \cite{lars87} operating under ExpGui platform \cite{toby01}; in general, only data from the highest-resolution backscattering detector bank were used.  All data for all samples were fit with a tetragonal model based on the $P4/nmm$ space group. In addition,the Fe$_{1+y}$Te data were modeled with $P21/m$ model below the phase transition temperature of $\sim65$~K.  For the FeSe sample, no phase transition was resolved.

\section{Results and Analysis}
\label{sec:raa}

\subsection{Magnetic Excitations}

\subsubsection{Data}

\begin{figure}[t]
\includegraphics[width=0.9\linewidth]{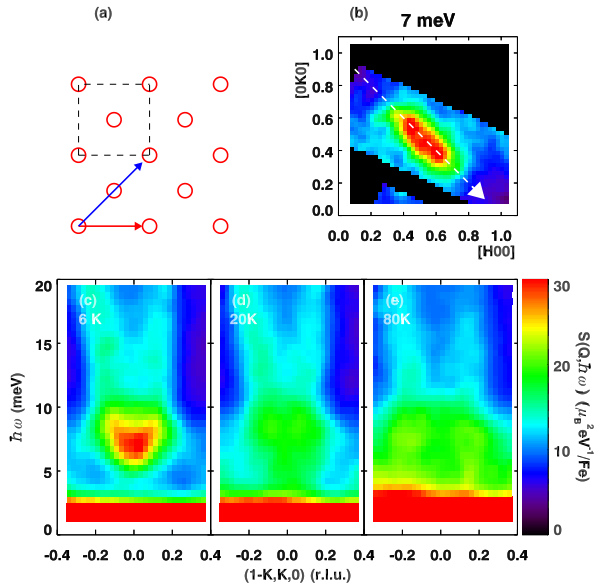}
\caption{(Color online) (a) The two-Fe unit cell used in the paper. The circles denote Fe atom positions. The red and blue arrows respectively denote the $[1,0]$ and $[1,1]$ directions in real space. (b) $Q_{AF}=(0.5,0.5)$, location of the spin-resonance in the $(H,K,0)$ plane. The dashed arrow denotes the transverse direction, along which we plot the inelastic magnetic neutron scattering for temperatures of (c) 6~K, (d) 20~K, and (e) 80~K, measured from the SC50 sample on SEQUOIA.
} \label{fig:1}
\end{figure}

To make contact with previous work, we begin with the measurements on the SC50 sample at SEQUOIA.   The magnetic excitation spectra about the stripe antiferromagnetic wave vector, ${\bf Q}_{\rm SAF}=(0.5,0.5)$, along the $[1,-1,0]$ direction is plotted for several temperatures in Fig.~\ref{fig:1}.
The spin gap of $\sim5$~meV is clearly visible in the 6~K  data.
The intensity near the bottom of the dispersion is clearly enhanced in the superconducting phase
and the magnetic excitations disperse outwards forming a ``U'' shape above this spin
resonance energy~\cite{Argyriou2009,Lee2010,Lis2010,Zhijun_2011}. In the normal state at $T=20$~K, the spin resonance fades away and broadens in energy and $Q$, so that intensity moves into the spin gap; nevertheless, the overall shape of the magnetic excitation spectrum does not appear to change significantly.

In contrast, significant changes are observed for excitations near the bottom of the dispersion when the sample is warmed to 80~K, far above $T_c$.  In Fig.~\ref{fig:1}~(e), one can see that the scattered signal has broadened considerably in $Q$ below 10~meV, and the bottom of the U-shaped dispersion appears to have split. This is consistent with the nominally incommensurate correlations previously observed in linear scans. Only for $\hbar\omega > 15$~meV does the spectrum remain relatively unchanged with temperature.


\begin{figure}[tb]
\includegraphics[width=0.8\linewidth]{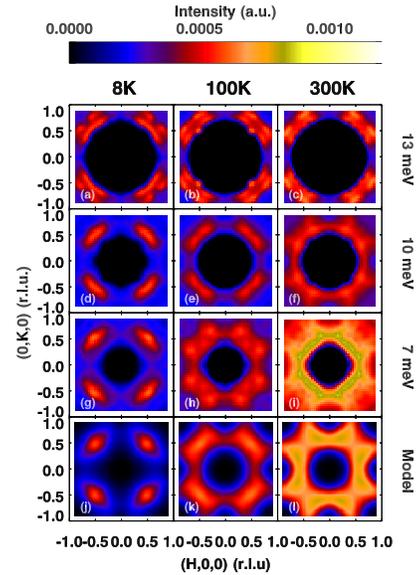}
\caption{(Color online) Inelastic magnetic neutron scattering from the SC70 sample measured on HYSPEC at energy transfers $\hbar\omega=$ 13~meV (a), (b), (c); 10~meV (d), (e), (f); and 7~meV (g), (h), (i).  The sample temperatures are 8~K (a), (d), (g); 100~K (b), (e), (h); and 300~K (c), (f), (i). All slices were taken with an energy width of 2~meV.  Measurements, covering approximately two quadrants, have been symmetrized to be 4-fold symmetric, consistent with sample symmetry.  Intensity scale is the same in all panels, but 13-meV data have been multiplied by 1.5 to improve visibility.  Black regions at the center of each panel are outside of the detector range. Panels (j), (k), (l) are model calculations simulating the 7-meV data, as described in the text, based on weakly correlated slanted UDUD spin plaquettes [see Fig~\ref{fig:scheme} (a) and (b)].  The wave-vectors for the AFM inter-plaquette correlations used in the calculation are (j) 100\% $Q_{SAF}$, (k) 50\% $Q_{SAF}$ and 50\% $Q_{DSAF}$, and (l) 100\% $Q_{DSAF}$.
} \label{fig:2}
\end{figure}

In order to understand the temperature-dependent changes in the low-energy scattering, however, we need to look at what is happening throughout the $(H,K,0)$ plane. For this, we turn to the measurements on the SC70 sample obtained at HYSPEC. Such constant-energy slices are plotted in Fig.~\ref{fig:2} for energy transfers of 7, 10, and 13 meV, and temperatures of 8, 100, and 300 K.
In the superconducting phase ($T=8~K$), the low energy ($\hbar\omega=7$~meV) magnetic excitations have ellipsoidal shapes centered on ${\bf Q}_{\rm SAF}$ positions, with the long axis oriented in  the transverse direction [Fig.~\ref{fig:2}~(g)].  The excitations at higher energies spread out along the transverse directions away from ${\bf Q}_{\rm SAF}$ [Fig.~\ref{fig:2}~(d),(a)], consistent with the dispersion shown in Fig.~\ref{fig:1}.  On warming to 100~K, the low-energy intensity maxima [Fig.~\ref{fig:2}~(g)] move away from ${\bf Q}_{\rm SAF}$, again consistent with Fig.~\ref{fig:1}.

At 300~K, the redistribution of low-energy signal in reciprocal space and an intensity enhancement are more pronounced. The intensities appear to form a continuous, though structured, ``squared'' ring about ${\bf Q}=(0,0)$. Corners of a ``squared'' ring going through the four (0.5,0) positions are reminiscent of 
the pattern observed
in FeTe$_{1-x}$S$_x$  \cite{Zaliznyak2015}.
In contrast, the temperature-induced changes in the intensity distribution are much less pronounced at 10~meV and are hardly noticeable at 13~meV, although the overall intensity increase is noticeable, similar to the trend observed in previous studies of FeTe and FeTe$_{1-x}$S$_x$ \cite{Zaliznyak2011,Zaliznyak2015}.

\begin{figure}[bt]
\includegraphics[width=0.8\linewidth]{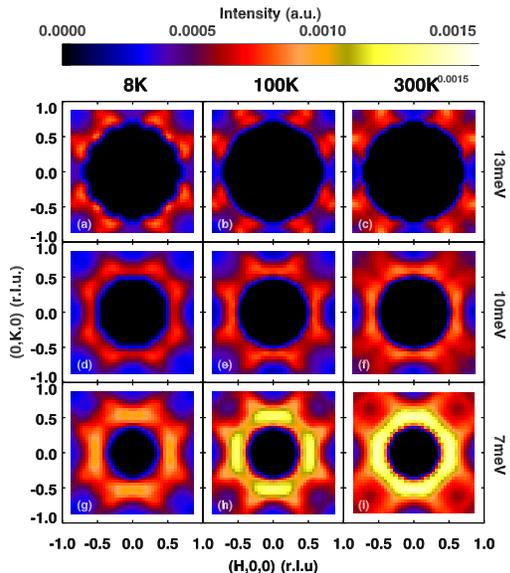}
\caption{(Color online) Inelastic magnetic neutron scattering measured at energy transfers $\hbar\omega=$ 13~meV (top row), 10~meV (2nd row from the top), and 7~meV (third row from the top) on HYSPEC. All slices were taken with an energy width of 2~meV. The sample used is the NSC45 sample. The temperatures for the measurements are 8~K, 100~K, 300~K, from left to right, respectively.
} \label{fig:3}
\end{figure}

Similar measurements were performed on the NSC45 sample, and the results are
shown in Fig.~\ref{fig:3}.  The data at 300~K are rather similar to the high-temperature data for SC70.  In contrast to SC70, however, cooling results in relatively little change to the scattering pattern at low energy, other than a reduction in intensity.   In fact, low-temperature measurements on triple-axis spectrometers have shown that broad elastic scattering centered at $(0.5,0,0.5)$ is present at low temperature for NSC45 but is absent for SC70 \footnote{Zhijun Xu, Jinsheng Wen, and Guangyong Xu (unpublished).}.


{To put these results in perspective, the data for the SC50 and SC70 samples are consistent with one another and with our previous results on good bulk superconducting samples~\cite{Xu2012PRL,Zhijun2014,Zhijun_2011}, though the latter results covered a more restricted part of {\bf Q}--$\omega$ space. Similarly, the data for the NSC45 sample are compatible with our previous measurements~\cite{Zhijun2014} on nonsuperconducting, Se-doped samples,  where excess Fe induces short-range antiferromagnetic order at low-temperature.}

\subsubsection{Modeling}

Several studies have suggested that the exchange couplings governing magnetic correlations in the iron chalcogenides are strongly frustrated, resulting in a variety of spin configurations having very similar free energies {\cite{Yinw2010,Fang2009,Glasbrenner2015,Yin2012,Luo2014}}. This frustration inhibits long-range magnetic order, and is qualitatively consistent with our observations of dynamical magnetic correlations with short correlation lengths.  Nevertheless, while the magnetic moments are clearly disordered, we find that they carry signatures of specific local spin configurations.

\begin{figure}[ht]
\includegraphics[clip,trim={2cm 0 0 0},width=1.3\linewidth]{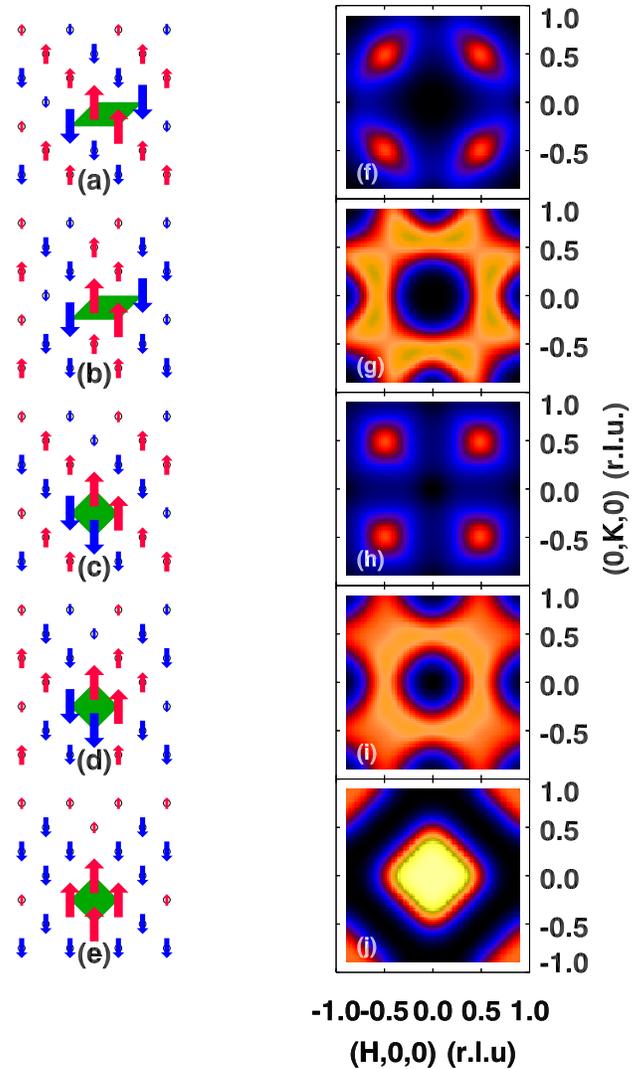}
\caption{(Color online) Schematics of the spin plaquettes in the weakly correlated spin-liquid model
described in the text. (a) A canted UDUD plaquette with modulation wave vector ${\bf Q}_{\rm SAF}$. (b) Canted UDUD plaquette with ${\bf Q}_{\rm DSAF}$.  (c) Square UDUD plaquette with ${\bf Q}_{\rm SAF}$. (d) Square UDUD plaquette with block antiferromagnetic correlations. (e) Square UUUU plaquette with block antiferromagnetic correlations. 
The frames (f) to (j) are model simulations described in the text, based on the liquid-like spin plaquette models in (a) to (e), respectively.
} \label{fig:scheme}
\end{figure}

Following Zaliznyak {\it et al.}\ \cite{Zaliznyak2015}, we consider models of static, short-range spin correlations that may represent a snapshot of the behavior for low-energy spin excitations---specifically, for our $\hbar\omega=7$~meV data.  Figure~\ref{fig:scheme} shows a variety of models and their 4-fold symmetrized Fourier transforms.  In each case, we start with a particular 4-spin plaquette; averaging over equivalent choices leads to the structure factor for the correlations.  We then choose a particular antiferromagnetic wave vector to describe inter-plaquette phasing, with an exponential decay of the correlations with distance between plaquettes.   For example, Fig.~\ref{fig:scheme}(a) and (b) use the same choice of plaquette (labeled up-down-up-down, or UDUD), but with longer-range correlations defined by ${\bf Q}_{\rm SAF}$ and ${\bf Q}_{\rm DSAF}$, respectively.  This results in dramatically different scattering patterns, as indicated by Fig.~\ref{fig:scheme}(f) and (g).  Figure~\ref{fig:scheme} also shows several other choices of plaquette and modulation wave vector.

Looking back at the 7-meV results for the SC70 sample in Fig.~\ref{fig:2}, we find that the data are quite similar to the calculations for the UDUD plaquette, but with the modulation wave vector changing with temperature, from ${\bf Q}_{\rm SAF}$ in the superconducting state to ${\bf Q}_{\rm DSAF}$ at room temperature.  At 100~K, a 50-50 mix of these models seems to apply.  The corresponding simulations are plotted in Fig.~\ref{fig:2}(j)--(l), for comparison with the data in (g)--(i). {The effective correlation length was chosen to be $\sim 0.5a$ in these cases, suggesting a highly disordered nature of the spin configuration that is consistent with a liquid phase. } For the NSC45 sample, as already mentioned, there appears to be no change in characteristic wave vector, ${\bf Q}_{\rm DSAF}$. 
Our data thus provide a direct probe of the spin-correlation wave-vector, in addition to the type of the local order. 
{In particular, we note that models with square plaquettes, such as Fig.~\ref{fig:scheme}(c) to (e), fail to reproduce important details of the data.}

\begin{figure}[t]
\includegraphics[width=0.9\linewidth]{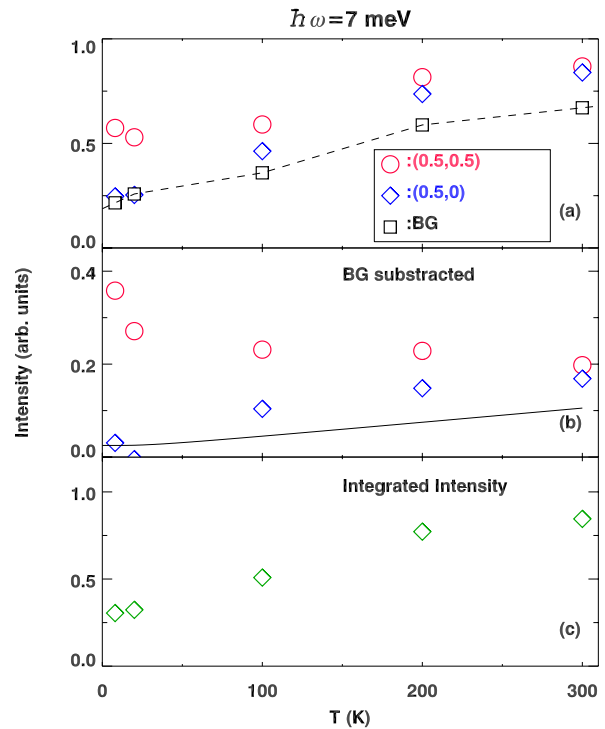}
\caption{(Color online) Inelastic neutron scattering intensity measured at $\hbar\omega=7$~meV with energy width $\delta E=1$~meV from the SC70 sample on HYSPEC. (a) Raw intensity
measured at ${\bf Q}_{\rm DSAF}=(0.5,0)$ (blue diamonds) and ${\bf Q}_{\rm SAF}=(0.5,0.5)$ (red circles). The numbers shown in the plot are
averaged intensities taken within a square region with $\delta H=0.1$ and $\delta K=0.1$ (r.l.u.) centered at the measurement wave-vectors.
The background is measured by averaging intensities
around $(1,0)$ and $(1,1)$, shown as black squares. (b) Background-subtracted intensities at ${\bf Q}_{\rm DSAF}$ (blue diamonds) and
${\bf Q}_{\rm SAF}$ (red circles). The solid line is the calculated detailed-balance factor. (c) Integrated intensity over the full Brillouin zone centered at (0,0).
} \label{fig:4}
\end{figure}

An alternative approach for analyzing the changing magnetic correlations is to simply compare the magnetic weight at the characteristic wave vectors ${\bf Q}_{\rm SAF}$ and ${\bf Q}_{\rm DSAF}$.
The raw and background (BG) subtracted intensities for these regions
at $\hbar\omega=7$~meV are plotted in Fig.~\ref{fig:4} (a) and (b). Here, the magnetic scattering intensity (BG subtracted) at ${\bf Q}_{\rm SAF} = (0.5,0.5)$ positions decreases with warming while that at ${\bf Q}_{\rm DSAF}=(0.5,0)$ increases.  For comparison, the solid line in
Fig.~\ref{fig:4}~(b) shows the detailed-balance factor $1/(1-e^{-\hbar\omega/k_BT})$, which would characterize the thermal evolution of collective excitations whose dynamical susceptibility does not depend on temperature for the range of temperature studied, {as one might expect in the case of a magnetically-ordered state. } 
The signal at $(0.5,0)$ clearly grows even faster than predicted by the detailed balance factor, which means that the dynamical magnetic susceptibility increases. 
The ratio of the signal at ${\bf Q}_{\rm SAF}$ to that at ${\bf Q}_{\rm DSAF}$ is plotted in Fig.~\ref{fig:comp}, where one can see that it evolves much like the nematicity from electroresistance measurements \cite{Kuo2015} and the inverse scattering rate of mobile charge carriers \cite{Homes2015}.

We also plot the background-corrected intensity integrated over the entire Brillouin zone for $\hbar\omega=7$~meV in Fig.~\ref{fig:4}~(c).  Due to kinematics, there is a region around ${\bf Q}=(0,0)$ that we cannot measure, so that we might miss some signal as the scattering spreads out with increasing temperature.
Nevertheless, from 10~K to 300~K, the $Q$-integrated low-energy spectral weight has increased by at least a factor of two.  At higher energy transfers, the  $Q$-range of our measurements becomes more limited and such integration over the entire Brillouin zone becomes unrealistic. Qualitatively, however, it is evident [see Fig.~\ref{fig:2}] that the increase of spectral weight with temperature also becomes much less pronounced at higher energy transfers.

\subsection{Temperature dependence of lattice tetragonality}

\begin{figure}[t]
\includegraphics[width=0.9\linewidth]{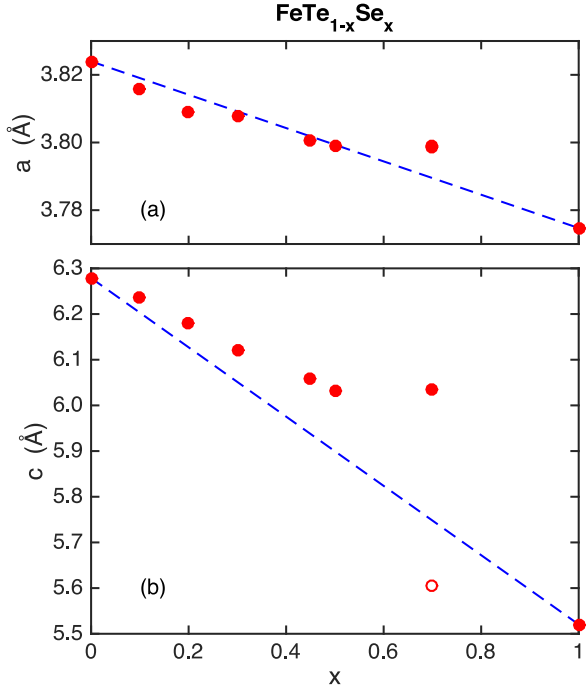}
\caption{(Color online) Lattice parameters $a$ (a), and $c$ (b), for FeTe$_{1-x}$Se$_x$ as a function of $x$, measured by neutron powder diffraction at 300 K on NOMAD.  Statistical uncertainties for $a$ and $c$ are smaller than the symbol size.  The dashed lines simply connect the points at $x=0$ and 1.
} \label{fig:acx}
\end{figure}

The observed thermal evolution of the magnetic correlations is {inconsistent with a model in which orbital hybridization and magnetic exchange couplings are independent of temperature.  Given the changes,} 
one might expect to see some sort of response in the lattice.  As already mentioned, symmetry-lowering structural transitions are common to Fe$_{1+y}$Te and FeSe.  For mixed compositions, the situation is complicated by the very different Fe-Te and Fe-Se bond lengths.  Scattering studies indicate that these bond lengths vary rather little \cite{Louca2010,Tegel2010}; as a consequence, scanning transmission electron microscopy \cite{Hu2011} and scanning tunneling microscopy \cite{He2011} studies provide evidence of segregation into Te-rich and Se-rich regions.  Such disorder may frustrate long-range ordering of distortions away from tetragonal symmetry; nevertheless, other behavior may survive.  Indeed, in our initial report of anomalous temperature-dependent changes of the magnetic correlations in FeTe$_{0.5}$Se$_{0.5}$ doped with Ni, we found an upturn in the $a$ lattice parameter at low temperature \cite{Xu2012PRL}.  Similar lattice behavior was reported earlier for FeTe$_{1-x}$Se$_x$ with $x=0.1$ and 0.2 by Martinelli {\it et al.} \cite{Martinelli}.  As a  result, we decided to take a more systematic look at the system with neutron powder diffraction measurements.

Figure~\ref{fig:acx} shows the room temperature values of the $a$ and $c$ lattice parameters as a function of $x$; here all samples have the tetragonal structure, with space group $P4/nmm$.    Note that the $c$ lattice parameter changes by 0.78~\AA\ (13\%) across this series, while $a$ changes by only 0.046~\AA\ (1.3\%).  This reflects the very different heights of the Te and Se ions relative to the Fe layer.  The $x=0.7$ sample is on the lower edge of the miscibility gap found by Fang {\it et al.} \cite{Fang2008PRB}.  We found evidence for two tetragonal phases, with the lattice parameters of the less dominant phase ($\sim 30$\% by volume) indicated by open circles ($a$ values are virtually identical). 
{Note that the second phase, with reduced $c$, is expected to be similar to FeSe, which is superconducting and lacks magnetic order.  The spin dynamics should be similar to that of the Te-doped superconducting phase, and no distinctive features were detected in the single-crystal inelastic measurements.}

\begin{figure}[t]
\includegraphics[width=0.9\linewidth]{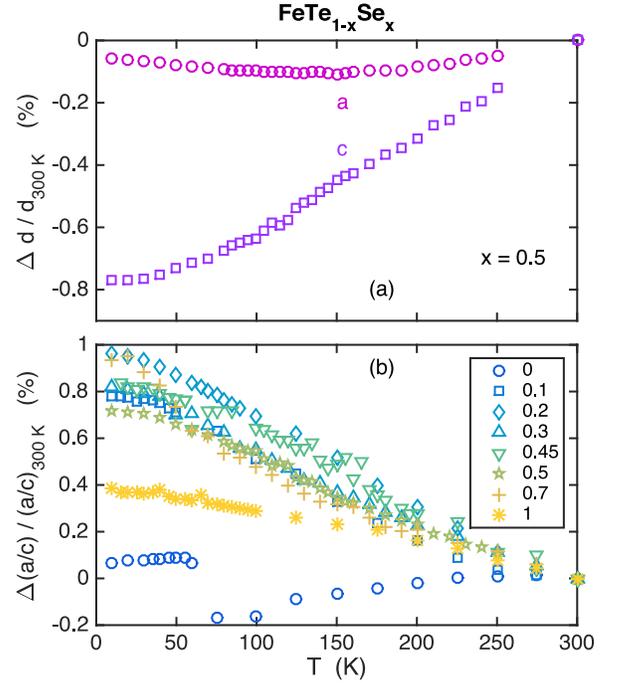}
\caption{(Color online) (a) Change in $a$ and $c$ lattice parameters, normalized to 300 K, as a function of temperature for the $x=0.50$ sample.  Statistical uncertainties are comparable to the symbol size. (b) Change in $a/c$, normalized to 300~K, as a function temperature for FeTe$_{1-x}$Se$_x$; the values of $x$ are noted in the symbol legend.  The average of in-plane lattice parameters was used for $a$ in the low-temperature phase of $x=0$.
} \label{fig:act}
\end{figure}

A representative example of the temperature dependence of the $a$ and $c$ lattice parameters for $x=0.5$ is plotted in Fig.~\ref{fig:act}(a).   As one can see, there is a distinct upturn in $a$ below $\sim150$~K, while the $c$ lattice parameter, if anything, appears to decrease a bit more rapidly in the same temperature range.  We observe very similar behavior for all samples in the range $0.1\le x\le0.7$.   To characterize this behavior, we have plotted the relative change in the $a/c$ ratio in Fig.~\ref{fig:act}(b).  To be specific, if $r=a/c$, then we plot $[r(T)-r(300\ {\rm K})]/r(300\ {\rm K})$.  In the case of $x=0$, we use the average in-plane lattice parameter in the low temperature phase; note that we did not resolve an orthorhombic phase in our $x=1$ sample.

To interpret this behavior, we note that the tetrahedral bond angle can be expressed as $\theta = \tan^{-1}(a/2zc)$, where $z$ is the relative coordinate of the chalcogenide ions.  Given the evidence for phase segregation in FeTe$_{1-x}$Se$_x$ \cite{Hu2011,He2011}, it should be reasonable to think about local Te-Fe-Te and Se-Fe-Se bond angles.  Regardless of the local $z$, the bond angle will move towards the ideal tetrahedral angle as $a/c$ increases.  The temperature dependence of $a/c$ shown in Fig.~\ref{fig:act}(b) indicates that bond angle increases toward the ideal on cooling, which implies a reduction of the crystal-field splitting \cite{Yin2011}, {and a change in hybridization}.  The relative change of the bond angle in each sample on cooling is relatively small; however, we believe it reflects a substantially larger change in {orbital content and occupancy of the electronic band structure}.  {We note that a recent ARPES study of FeTe$_{1-x}$Se$_x$ with $x=0.44$ has found a significant growth on cooling for spectral weight of the $d_{xy}$ band near the Fermi level \cite{Yi2015}.}

\section{Summary and Discussion}
\label{sec:disc}

By mapping the magnetic scattering over the entire $(H,K,0)$ plane of reciprocal space, we have shown that the characteristic wave vector of the low-energy spin correlations shifts from ${\bf Q}_{\rm DSAF}$ to ${\bf Q}_{\rm SAF}$ on cooling in superconducting FeTe$_{1-x}$Se$_x$.  The ratio of the magnetic signal at the latter point to the former grows at low temperature much like the nematic response of elastoresistance measurements \cite{Kuo2015} and the inverse scattering rate of the mobile carriers \cite{Homes2015}, as shown in Fig.~\ref{fig:comp}. In a sample rendered non-superconducting by inclusion of excess Fe, the magnetic wave vector is  ${\bf Q}_{\rm DSAF}$ and shows no thermal shift. In all samples studied, local correlations are consistent with antiferromagnetic UDUD plaquettes having C$_2$ local symmetry indicative of nematicity, in agreement with the study of Ref. \cite{Zaliznyak2015}, where such correlations were found to develop with doping towards superconductivity. We thus conclude that the change in the wave vector which describes propagation of magnetic correlations from ${\bf Q}_{\rm DSAF}$ to ${\bf Q}_{\rm SAF}$ is a further necessary condition for superconductivity in 11 iron chalcogenides.

{In our superconducting samples, the modeling of {\bf Q} dependence of the low-energy magnetic scattering suggests local rotational symmetry breaking at all temperatures.  However, it is the temperature-dependent change in characteristic magnetic wave vector that seems to qualitatively correlate with the growth in the nematic response of the elastoresistance measurements \cite{Kuo2015}, as indicated in Fig.~\ref{fig:comp}. }
The variation in antiferromagnetic wave vector implies a relative change among the exchange couplings over various Fe-Fe neighbor distances.  
A likely cause of this change is a temperature dependent variation of the orbital overlaps, the corresponding hybridization, as well as the variation in the occupancies of Fe $d$ levels.  {The ARPES evidence for local splitting of $xz$ and $yz$ bands \cite{Johnson2015}, together with the temperature-dependent nematic response \cite{Kuo2015}, supports this sort of variation.} 
Our evidence for the thermal variation of the tetrahedral bond angle indicates a modification of the splitting, and hence {the orbital content and the occupancy of the $xy$ and $xz/yz$ based bands that cross the Fermi level.} 
 {Regarding the question of what drives the nematic response, we can conclude that it is {\it not} an approach to magnetic order; local orbital order is a more likely suspect.  Nevertheless, it is clear that the magnetic, orbital, and lattice degrees of freedom are strongly entangled.}

We have also observed a reduction of low-energy magnetic spectral weight on cooling.  This is consistent with nuclear magnetic resonance results for an $x=0.5$ sample in which the quantity $1/T_1T$, where $1/T_1$ is the spin-lattice relaxation rate measured at the Te site, decreases as the temperature is reduced \cite{Shimizu2009}.  This loss of magnetic susceptibility is correlated with a growth in electronic conductivity \cite{Homes2015,Sun2015} and a crossover from incoherence to coherence.  This correlation parallels the more extreme changes observed in Fe$_{1+y}$Te \cite{Zaliznyak2011,Fobes2014}.   Theoretically, a competition between antiferromagnetic correlations and conductivity is expected \cite{Haule2009,Tam2015}.  The same electrons must contribute to the magnetic moments, influenced by Coulomb and Hund's interactions, and to electronic conductivity, minimizing kinetic energy.  This balance adjusts on cooling, with changing hybridization, and this competition likely plays an important role in determining the superconducting state.

{Altogether, there is evidence for temperature-correlated changes in characteristic magnetic wave vector, nematicity, electronic coherence, magnetic spectral weight, and tetrahedral bond angle. {It would be surprising if there is not an underlying connection why all of these happen together, and the temperature-dependent orbital hybridization and the orbital-selective electronic coherence provide a plausible connection.} These relationships certainly deserve further study.  }

\section{Acknowledgment}
We are grateful for stimulating discussions with Wei Ku, Ian Fisher, Adriana Moreo, Elbio Dagotto, and Ming Yi. The work at Brookhaven National Laboratory was supported by the Office of Basic Energy Sciences, U.S.\ Department of Energy, under Contract No.\ DE-SC00112704. Z.J.X.\ and R.J.B.\ are also supported by the Office of Basic Energy Sciences, U.S.\ Department of Energy through Contract No.\ DE-AC02-05CH11231. Research at Oak Ridge National Laboratory was sponsored by the Division of Scientific User Facilities of the same Office. The work at Nanjing University was supported by NSFC No.11374143, and NCET-13-0282.

\end{document}